

\input phyzzx   


%
\catcode`\@=11 
\paperfootline={\hss\iffrontpage\else\ifp@genum\tenrm
    -- \folio\ --\hss\fi\fi}
\def\titlestyle#1{\par\begingroup \titleparagraphs
     \iftwelv@\fourteenpoint\fourteenbf\else\twelvepoint\twelvebf\fi
   \noindent #1\par\endgroup }
\def\eqnalign{\eqname}
\def\GENITEM#1;#2{\par \hangafter=0 \hangindent=#1
    \Textindent{#2}\ignorespaces}
\def\papersize{\hsize=35pc \vsize=52pc \hoffset=0.5pc \voffset=0.8pc
   \advance\hoffset by\HOFFSET \advance\voffset by\VOFFSET
   \pagebottomfiller=0pc
   \skip\footins=\bigskipamount \normalspace }
\papers  
\def\address#1{\par\kern 5pt \titlestyle{\twelvepoint\sl #1}}
\def\abstract{\par\dimen@=\prevdepth \hrule height\z@ \prevdepth=\dimen@
   \vskip\frontpageskip\centerline{%
	\iftwelv@\fourteencp\else\twelvecp\fi Abstract}\vskip\headskip }
\newif\ifYITP \YITPtrue
\font\fourteenmib =cmmib10 scaled\magstep2    \skewchar\fourteenmib='177
\font\elevenmib   =cmmib10 scaled\magstephalf   \skewchar\elevenmib='177
\def\YITPmark{\hbox{\fourteenmib YITP\hskip0.2cm
        \elevenmib Uji\hskip0.15cm Research\hskip0.15cm Center\hfill}}
\def\titlepage{\FRONTPAGE\papers\ifPhysRev\PH@SR@V\fi
    \ifYITP\null\vskip-1.70cm\YITPmark\vskip0.6cm\fi 
   \ifp@bblock\p@bblock \else\hrule height\z@ \rel@x \fi }
\catcode`\@=12 
%
%
\newbox\leftpage \newdimen\fullhsize
\catcode`\@=11 
\ifx\Landscapeout\Undefined\message{(This is unreduced.}
\else
\tolerance=1000\hfuzz=2pt
\message{(This is the preprint format.} \let\l@r=L
\def\refout{\par\penalty-400\vskip\chapterskip
   \spacecheck\referenceminspace
   \ifreferenceopen \Closeout\referencewrite \referenceopenfalse \fi
   \line{\twelverm\hfil REFERENCES\hfil}\vskip\headskip
   \input \jobname.refs
   }
\def\figout{\par\penalty-400
   \vskip\chapterskip\spacecheck\referenceminspace
   \iffigureopen \Closeout\figurewrite \figureopenfalse \fi
   \line{\twelverm\hfil FIGURE CAPTIONS\hfil}\vskip\headskip
   \input \jobname.figs
   }
\magnification=1000\baselineskip=15pt plus 0.2pt minus 0.1pt\vsize=7truein
\fullhsize=11truein
\def\papersize{\hsize=4.9truein\vsize=7truein\hoffset=-0.3in\voffset=-0.5in
	\pagebottomfiller=0pc \skip\footins=\bigskipamount  }
\paperfootline={\hss\iffrontpage\else\ifp@genum\ninerm
	-- \the\count0\ -- \hss\fi\fi}
\font\twelvemib   =cmmib10 scaled\magstep1	    \skewchar\twelvemib='177
\font\tenmib      =cmmib10			    \skewchar\tenmib='177
\def\YITPmark{\hbox{\twelvemib YITP\hskip0.2cm
        \tenmib Uji\hskip0.15cm Research\hskip0.15cm Center\hfill}}
\def\titlepage{\FRONTPAGE\papers\ifPhysRev\PH@SR@V\fi
    \ifYITP\null\vskip-1cm\YITPmark\vskip0.6cm\fi 
   \ifp@bblock\p@bblock \else\hrule height\z@ \rel@x \fi }
\output={\almostshipout{\leftline{\vbox{\pagebody\makefootline}}}
	\advancepageno}
\def\almostshipout#1{\if L\l@r \count1=1 \message{[\the\count0.\the\count1]}
      \global\setbox\leftpage=#1 \global\let\l@r=R
 \else \count1=2
  \shipout\vbox{{\hsize\fullhsize\makeheadline}
      \hbox to\fullhsize{\box\leftpage\hfil#1\hss}}  \global\let\l@r=L\fi}
\tenpoint
\fi
\catcode`\@=12 
%
\catcode`\@=11 
\let\rel@x=\relax
\let\n@expand=\relax
\def\pr@tect{\let\n@expand=\noexpand}
\let\protect=\pr@tect
\let\gl@bal=\global
%
%
\newtoks\t@a \newtoks\t@b \newtoks\next@a
\newcount\num@i \newcount\num@j \newcount\num@k
\newcount\num@l \newcount\num@m \newcount\num@n
\long\def\l@append#1\to#2{\t@a={\\{#1}}\t@b=\expandafter{#2}%
                         \edef#2{\the\t@a\the\t@b}}
\long\def\r@append#1\to#2{\t@a={\\{#1}}\t@b=\expandafter{#2}%
                         \edef#2{\the\t@b\the\t@a}}
\def\l@op#1\to#2{\expandafter\l@opoff#1\l@opoff#1#2}
\long\def\l@opoff\\#1#2\l@opoff#3#4{\def#4{#1}\def#3{#2}}
\newif\ifnum@loop \newif\ifnum@first \newif\ifnum@last
\def\sort@@#1{\num@firsttrue\num@lasttrue\sort@t#1}
\def\sort@t#1{\pop@@#1\to\num@i\rel@x
            \ifnum\num@i=0 \num@lastfalse\let\next@a\rel@x%
            \else\num@looptrue%
                 \loop\pop@@#1\to\num@j\rel@x
                    \ifnum\num@j=0 \num@loopfalse%
                    \else\ifnum\num@i>\num@j%
                         \num@k=\num@j\num@j=\num@i\num@i=\num@k%
                         \fi%
                    \fi%
                    \push@\num@j\to#1%
                  \ifnum@loop\repeat%
                  \let\next@a\sort@t%
            \fi%
            \print@num%
            \next@a#1}
\def\print@num{%
              \ifnum@first%
                 \num@firstfalse\num@n=\num@i\number\num@i%
              \else%
                 \num@m=\num@i\advance\num@m by-\num@l%
                 \ifcase\num@m\message{%
                   *** WARNING *** Reference number %
                   [\the\num@i] appears twice or more!}%
                 \or\rel@x%
                 \else\num@m=\num@l\advance\num@m by-\num@n%
                    \ifcase\num@m\rel@x%
                    \or,\number\num@l%
                    \else-\number\num@l%
                    \fi%
                    \ifnum@last\num@n=\num@i,\number\num@i\fi%
                 \fi%
              \fi%
              \num@l=\num@i%
              }
\def\pop@@#1\to#2{\l@op#1\to\z@@#2=\z@@}
\def\push@#1\to#2{\edef\z@@{\the#1}\expandafter\r@append\z@@\to#2}
\def\append@cs#1=#2#3{\xdef#1{\csname%
                    \expandafter\g@bble\string#2#3\endcsname}}
\def\g@bble#1{}
\def\if@first@use#1{\expandafter\ifx\csname\expandafter%
                              \g@bble\string#1text\endcsname\relax}
\def\keep@ref#1#2{\def#1{0}\append@cs\y@@=#1{text}\expandafter\def\y@@{#2}}
\def\keepref#1#2{\if@first@use#1\keep@ref#1{#2}%
                 \else\message{%
                    \string#1 is redefined by \string\keepref! %
                    The result will be .... what can I say!!}%
                 \fi}
\def\Null{0}
\def\get@ref#1#2{\def#2{\string#1text}}
\def\findref@f#1{%
                \ifx#1\Null \get@ref#1\text@cc\R@F#1{{\text@cc}}%
                \else\rel@x\fi}
\def\findref#1{\findref@f#1\ref@mark{#1}}
\def\find@rs#1{\ifx#1\endrefs \let\next=\rel@x%
              \else\findref@f#1\r@append#1\to\void@@%
                    \let\next=\find@rs \fi \next}
\def\findrefs#1\endrefs{\def\void@@{}%
                    \find@rs#1\endrefs\r@append{0}\to\void@@%
                    \ref@mark{{\sort@@\void@@}}}
\let\endrefs=\rel@x
\def\ref@mark#1#2{\if#2,\rlap#2\refmark#1%
                  \else\if#2.\rlap#2\refmark#1%
                   \else\refmark{{#1}}#2\fi\fi}
\def\NPref@mark#1#2{\if#2,\NPrefmark#1,%
                  \else\if#2.\NPrefmark#1.%
                   \else\thinspace\NPrefmark{{#1}} #2\fi\fi}
\def\R@F#1{\REFNUM #1\R@F@WRITE}
\def\NPrefitem#1{\r@fitem{[#1]}}
\def\NPrefs{\let\refmark=\NPrefmark \let\refitem=\NPrefitem%
            \let\ref@mark=\NPref@mark}
\def\PRrefs{\let\refmark=\PRrefmark}
\def\R@F@WRITE#1{\ifreferenceopen\else\gl@bal\referenceopentrue%
     \immediate\openout\referencewrite=\jobname.refs%
     \toks@={\begingroup \refoutspecials}%
     \immediate\write\referencewrite{\the\toks@}\fi%
    \immediate\write\referencewrite{\noexpand\refitem%
                                    {\the\referencecount}}%
    \immediate\write\referencewrite#1}
\catcode`\@=12 
\ifx\epsfbox\Undefined\def\physfig#1#2#3{\FIGURE#1{#2}
	\goodbreak\midinsert
	\vskip1cm
           \par\vskip0.5cm{\hbox{\centerline{%
           \vbox{\hrule height0.5pt%
           \hbox{\vrule width0.5pt\hskip4pt%
           \vbox{\vskip5pt\hbox{Fig.~\the\figurecount}\vskip4pt}%
           \hskip4pt\vrule width0.5pt}%
           \hrule height0.5pt}}}}
	\vskip1cm
	\bigskip\endinsert}
\else\def\physfig#1#2#3{\FIGURE#1{#2}
	\goodbreak\midinsert\centerline{#3}
	\bigskip\centerline{\vbox{
	\advance\hsize by -30mm\noindent
	{\bf Fig.~\the\figurecount:} #2}}\bigskip\endinsert}
\fi
\Pubnum={YITP/U-92-37\cr hep-th/9301020}
\date={December 1992}

\titlepage
\title{ On the Mass of Two Dimensional Quantum Black Hole }

\author{Tsukasa TADA
\foot{E-mail address: tada@yisun1.yukawa.kyoto-u.ac.jp}
\foot{Soryuushi Shougakukai Fellow}
{\tenrm and} Shozo UEHARA
\foot{E-mail address: uehara@yisun1.yukawa.kyoto-u.ac.jp}
\foot{Work partially supported by Grant-in-Aid for Scientific Research
	of the Ministry of Education, Science and Culture No.02302020}
}

\address{Uji Research Center \break
               Yukawa Institute for Theoretical Physics\break
               Kyoto University,~Uji 611,~Japan}

\abstract{ For the two dimensional dilaton-coupled quantum gravity
model, we give the local black hole mass, which is an analogue of
what was first introduced by Fischler, Morgan and Polchinski in the
four dimensional gravitational systems.
We analyze the original CGHS model with this local mass and find that
the local mass is decreasing in the future direction on the matter
shock-wave line, while it stays constant at past null infinity.  }

\endpage
\sequentialequations
\NPrefs
\keepref\hawking{S.W.~Hawking, Comm. Math. Phys. {\bf 43} (1975) 199.}
\keepref\cghs{C.G.~Callan, S.B.~Giddings, J.A.~Harvey and
	A.~Strominger, Phys. Rev. {\bf D45} (1992) R1005.}
\keepref\bs{T.~Banks, A.~Dabholkar, M.R.~Douglas and
	M.~O'Loughlin, Phys. Rev. {\bf D45} (1992) 3607.}
\keepref\susskind{J.G.~Russo. L.~Susskind and L.~Thorlacius,
		Phys. Let. {\bf B292} (1992) 13;
		preprint SU-ITP-92-17 (1992);
		preprint UTTG-19-92 (1992).}
\keepref\bc{A.~Biral and C.~Callan, preprint
		PUPT-1320,hepth@xxx/9205089 (1992).}
\keepref\strominger{A.~Strominger, preprint UCSBTH-92-18,
	hepth@xxx/9205028 (1992).}
\keepref\dealwis{S.P.~de~Alwis, preprint COLO-HEP-280,
	hep@xxx9205069 (1992); preprint COLO-HEP-284, hep@xxx9206020 (1992).}
\keepref\hamada{K.~Hamada, preprint UT-Komaba 92-7 (1992).}
\keepref\fmp{W.~Fischler, D.~Morgan and J.~Polchinski, Phys. Rev.
	{\bf D42} (1990) 4042.}
\keepref\tomimatsu{A.~Tomimatsu, Phys. Lett. {\bf B289} (1992) 283.}


	Recently two dimensional black hole physics has been
attracting much interest in studying the evaporation of black holes.
It raised some hope of solving problems associated with the Hawking
radiation\findref\hawking.
Callan, Giddings, Harvey and Strominger\findref\cghs studied the
string-inspired two dimensional toy model (CGHS model).
They presented the general solutions at the classical level and showed
that their model has a solution having a black hole which is formed by
the matter shock wave. And also they discussed about the evaporation
of the black hole at the quantum level. Subsequently many features of
the CGHS model and its modified versions have been vigorously studied
by many
people\findrefs\bs\susskind\hamada\bc\dealwis\strominger\endrefs.

	In this paper we shall study the evaporation of the black hole
concentrating on the mass of the black hole in the original CGHS
model.  First we will define a local function which gives the mass of
a black hole at the classical level. And then we will analyze the
behavior of the mass function at the quantum level.

The CGHS model without the conformal matter fields is given by
$$
S = {1 \over 2 \pi}\int d^2x \sqrt{-g}\ e^{-2\phi}
	\left( R + 4 (\nabla\phi)^2+ 4 \lambda^2 \right), \eqn\ichi
$$
where $\phi$ is a dilaton field and $\lambda$ is a constant.
Following the Arnowitt-Deser-Misner (ADM) formulation,
the two dimensional metric $g$ is written by
$$
	g_{ab}=\pmatrix{ -N_0^2+{N_1^2 \over \gamma}& N_1 \cr
	N_1&\gamma\cr}~,\eqn\ni
$$
where $N_0$ ($N_1$) is known as lapse (shift) and $\gamma$ stands for a
dynamical degree of freedom of gravitational sector.  $\gamma$ and the
dilaton $\phi$ are the dynamical degrees of freedom in the present
model Eq.\ichi.
	In terms of the canonical variable, that is,  $\gamma,\phi$
and their conjugate momenta, $\pi_\gamma, \pi_\phi$ , the above action
is rewritten as,
$$
S=\int \pi_\phi {\dot \phi}+ \pi_\gamma{\dot \gamma}
-N_0 {\cal H}_0 -N_1{\cal H}_1~.\eqn\san
$$
Here the dot stands for time derivative and ${\cal H}_0$
(${\cal H}_1$) is the generator of time (space) reparametrization;
$$
\eqalign{
{\cal H}_0 &= 2\,\sqrt{\gamma}\,e^{2 \phi}\,\pi_{\gamma}\,
	\pi_{\phi} + 4\,\gamma\,\sqrt{\gamma}\,e^{2
	\phi}\,\pi_{\gamma}^2 -
	{ e^{-2\phi} \left(\phi	'\right)^2\over{ \sqrt{\gamma}}}
	- \sqrt{\gamma}\,e^{-2\phi}\,\lambda^2~,\cr
{\cal H}_1 &= - {\gamma '\over \gamma}\,\pi_{\gamma} +
	{\phi ' \over\gamma}\,\pi_{\phi}~,\cr}\eqn\sana
$$
where the prime denotes the spatial derivative.
The lapse and shift are Lagrange multipliers, which leads to the
constraint equations,
$$
\eqalign{
	{\cal H}_0 (\pi_\phi, \phi, \pi_\gamma, \gamma) &=0, \cr
	{\cal H}_1 (\pi_\phi, \phi, \pi_\gamma, \gamma) &=0. \cr
}\eqn\yon
$$

	We find that the combination of the above constraints,
$$
\Phi \equiv {2\phi' \over \lambda\sqrt{\gamma}} \times {\cal H}_0
	- {4\gamma\,\pi_\gamma\,e^{2 \phi} \over \lambda}
	\times {\cal H}_1~,\eqn\go
$$
becomes the total derivative of a local function,
$$
\eqalignno{
\Phi &= {\cal M}'  = 0~,&\eqnalign{\rokua}\cr
{\cal M} &\equiv {4\gamma\,\pi_\gamma^2 \over \lambda}\,e^{2 \phi}
	- {(\phi')^2 \over \lambda\gamma}\,e^{-2 \phi}
	+ \lambda\,e^{-2\phi}~.&\eqnalign{\rokub} \cr
}
$$
This quantity ${\cal M}$ is the two dimensional version of
a local mass which is first introduced by Fischler Morgan and
Polchinski\findref\fmp in the study of a spherically symmetric
four dimensional gravitational system.
It also plays an important role in studying the evaporation of the four
dimensional black hole in Ref.\findref\tomimatsu.
199z
	The local mass function \rokub\  becomes
$$
{\cal M}= {1\over \lambda} \left[
	4 e^{-2\rho}e^{-2\phi} \partial_+ \phi \partial_- \phi
	+ \lambda^2 e^{-2\phi} 	\right]\eqn\hachi
$$
in the conformal gauge;
$$
g_{+-}= - {1 \over 2}e^{2\rho},\quad g_{++}=g_{--}=0,\eqn\nana
$$
where the light-cone coordinates are $x^{\pm} = x^0 \pm x^1$.
Plugging the classical static black hole solution of mass $M$,
$$
ds^2= e^{2 \rho} dx^+dx^- = {dx^+ dx^- \over {M \over
\lambda}-\lambda^2 x^+ x^-}, \qquad \phi=\rho,\eqn\kyu
$$
into the above mass function, we find
$$
{\cal M}(x^+,x^-)=M.\eqn\ju
$$
Furthermore, in Ref.[\cghs] there presented an example of the
formation of the black hole by the shock wave of a conformal matter
field $f$ traveling in the $x^-$ direction at $x^+ = x^+_0$.
The stress tensor of the matter is given by
$$
{1\over 2} \partial_+ f\,\partial_+ f = {M\over \lambda x^+_0}
	\,\delta(x^+-x^+_0),\eqn\jichi
$$
where $M$ is a parameter representing the magnitude of the shock-wave,
which is shown to be the mass of a black hole.
Then the classical solution is
$$
\eqalign{
	ds^2 &=e^{2 \rho} dx^+ dx^-={dx^+ dx^- \over -
	{M \over \lambda x^+_0} (x^+-x^+_0)\theta(x^+-x^+_0)
	- \lambda^2 x^+ x^-},\cr
	\rho &= \phi.\cr}\eqn\jni
$$
Calculating the local mass function for this geometry, we get
$$
{\cal M} (x^+,x^-)= M \theta(x^+-x^+_0).\eqn\jsan
$$
The value becomes zero in the Linear Dilaton Vacuum (LDV) region while
beyond the matter shock-wave line it becomes $M$ as expected.

	So far we have seen that the mass function gives the mass of
the black hole at the classical level.  Eq.\jsan\ means that the
evaporation of the black hole does not occur classically since the
mass of the black hole does not vanish towards the future null
infinity.

	Now we consider how the mass function behaves at the quantum
level. At the one-loop level, the quantum corrections are the
contributions of the conformal anomaly of the matter fields and that
from the gravitational sector.  We incorporate the quantum effect
through including the following term which comes from the trace
anomaly into the action,
$$
	\kappa\,\partial_+ \partial_- \rho ,\eqn\jyon
$$
where $\kappa$ depends on the number of the matter fields and here we
assume that it is a large positive number. Then we have the following
equations at this level,
$$
\eqalignno{
{}&{}\partial_+ \partial_- \phi = \left( 1- {\kappa\over 2}\,
	e^{2\phi} \right) \partial_+ \partial_- \rho~,&\eqnalign{\jgoa}\cr
{}&{}2\left(1-\kappa\,e^{2\phi}\right)\,\partial_+\partial_-\phi
	- 4\left(1-{\kappa\over 2}\,e^{2\phi}\right)\,\partial_+\phi
	\partial_-\phi - \left(1-{\kappa\over 2}\,e^{2\phi}\right)\,
	\lambda^2\,e^{2\rho} = 0~.\eqnalign{\jgob}\cr}
$$

	Once incorporating the quantum effect, one finds that the
model is no longer exactly solvable. Many features in classical theory
become different. For example, $\phi$ does not stay equal to $\rho$,
and the analysis breaks down due to the singularity at some value of
$\phi$, etc. On the other hand, Linear Dilaton Vacuum (LDV) is still a
solution of quantum system, that is, LDV is stable in this quantum
theory.

	Then we will analyze on a narrow region above the matter
shock-wave line $x^+ = x^+_0$ and also the past null infinity region.
We assume that the fields $\phi$ and $\rho$ take the classical values
on the line, which guarantees that the solution approaches the
classical one asymptotically.

	 Since the mass function is local we might see the
spacetime point where the evaporation completes. However the quantum
effect will be large at that point and since we are restricted to the
one-loop level, the point may be beyond our scope in the present
paper.

	Now we shall see the behavior of the mass function with the
quantum correction.
	On the matter shock-wave line, the mass function\hachi\ can be
obtained if one knows $\partial_+\phi(x_0^+,x^-)$. It can be calculated
from Eq.\jgob\findref\susskind,
$$
\partial_+\phi(x_0^+,x^-)= -{1\over 2x_0^+} +
	{M\over{2 \lambda x_0^+}}\,{1\over{\sqrt{w}\,
	\sqrt{w-\kappa}}}\ ,\eqn\jroku
$$
where
$$
	w \equiv -\lambda^2 x_0^+ x^-~.\eqn\jrokua
$$
Then we have
$$
{\cal M}\big|_{x^+ = x_0^+}=M \sqrt{w\over{w-\kappa}}~.\eqn\jnana
$$
${\cal M}$ diverges at $w = \kappa$ where the two
dimensional curvature is singular\findrefs\bs\susskind\endrefs.
The quantum effect is very large at that singular point and
our analysis has already been broken down there. On the other hand, at
the apparent horizon on the matter shock-wave line\findref\susskind,
$$
x^-_{AH}=-\sqrt{\left(M\over{\lambda^3 x_0^+}\right)^2 +
	\left(\kappa\over{2 \lambda^2 x_0^+}\right)^2}
	- {\kappa\over{2\lambda^2 x_0^+}}~,\eqn\jhachi
$$
or
$$
w_{AH}=\sqrt{\left( M\over \lambda\right)^2 +
	\left(\kappa\over 2\right)^2}
	+ {\kappa\over 2}~.\eqn\jhachia
$$
${\cal M}$ is finite and towards the past null infinity
$x^- \rightarrow -\infty$ it decreases to $M$.

	Similarly we can calculate $\partial_+\phi(x_0^+
+\epsilon,x^-)$ with small $\epsilon$ and $\partial_+\rho(x_0^+,x^-)$
from Eqs.\jgoa\ and \jgob.
And hence we obtain the mass function at $x^+=x^+_0+\epsilon$,
$$
\eqalign{
{\cal M}\big\vert_{x^+=x^+_0 + \epsilon} &=
	{M\,\sqrt{w}\over\sqrt{w-\kappa}} \cr
{}& +{M \epsilon\over{x^+_0}}\Biggl[
	{M\,w\over{\lambda\,(w-\kappa)^2}}
	-{5\,\kappa\,\sqrt{w}\over{4\,(w-\kappa)^{3\over2}}}
	-{2\,M\,w\over{\lambda\,\kappa\,(w-\kappa)}}\cr
{}& \hskip15mm+ {2\,M\,\sqrt{w}\over{\lambda\kappa\,\sqrt{w-\kappa}}}
	+{\sqrt{w}\over{4\,\sqrt{w-\kappa}}}\,\log{w\over{w-\kappa}}
	\Biggr]~.\cr}
	\eqn\jkyu
$$

	And if we assume that the quantum fluctuation becomes small
towards the past null infinity and the fields become almost classical
there, we can calculate the mass function similarly as,
$$
{\cal M}\big\vert_{x^- \rightarrow -\infty} = M~.\eqn\njyu
$$
Furthermore, the derivative of the mass function with respect to $x^-$
is obtained by
$$
\partial_-{\cal M}\big\vert_{x^- \rightarrow -\infty}= 0~.\eqn\njichi
$$
The mass function is shown in figure 1.

\physfig\fichi{The local mass function Eq.\rokub. The region painted
black is over the singularity. The mass function in the blank region
is beyond our present analysis.}{\epsfbox{dilfig1.eps}}

The behavior of the mass function has the following features:
i) it decreases along the negative $x^-$ direction on the
matter shock-wave line; ii) it decreases in the $x^+$ direction
on the matter shock-wave line $(x^+_0, - \infty < x^- < x^-_{AH})$;
iii) at $x^- \rightarrow -\infty$ the mass function is constant and its
derivative with respect to $x^-$ is zero.

	We have defined the local mass function \hachi\ which gives
the mass of the black hole in the two dimensional dilaton gravity
system at the classical level.
	The evaporation of the black hole at the quantum level implies
that the mass function decreases to zero. We found that the mass
function decreases in the $x^+$ direction on the matter shock wave
line, however it was not detected that the mass function becomes below
the classical value due to the quantum effect in our analysis. Further
analysis towards the future null infinity or along the apparent
horizon is necessary.

\ack

One of the authors (T.T) would like to thank Soryuushi Shougakukai for
the financial support and M. Sasaki for discussions.

\refout
\ifx\epsfbox\Undefined\figout\fi
\bye